\documentstyle[pra,aps,multicol,eqsecnum,epsfig]{revtex}

\newcommand\be{\begin{eqnarray}}
\newcommand\ee{\end{eqnarray}}
\newcommand\ba{\begin{array}}
\newcommand\ea{\end{array}}
\def\r{\rangle}
\def\l{\langle}
\def\T{{\rm Tr}}

\def\U{{\bf U}}

\def\cE{{\cal E}}

\begin{document}
\title{Role of entanglement and correlations in  dense coding}
\author{M\'ario Ziman$^{1}$ and Vladim\'\i r Bu\v zek$^{1,2}$}
\address{
${}^{1}$ Research Center for Quantum Information, 
Slovak Academy of Sciences,
D\'ubravsk\'a cesta 9, 
84228 Bratislava, Slovakia\\
${}^{2}$ Faculty of Informatics, Masaryk University, Botanick\'a 68a,
602 00 Brno, Czech Republic
}
\date{May 10, 2002}
\maketitle
\begin{abstract}
Usually it is assumed that quantum dense coding is due to quantum
entanglement between two parties. 
We  show that this phenomenon has its origin in
{\em correlations} between two parties rather than simply in entanglement.
In order to justify our argument we  considered that 
Alice has a  qubit  in the state
$\varrho=\frac{1}{2}\openone+\vec{n}\cdot\vec{\sigma}$
and we evaluate a capacity of the noiseless channel for two cases: 
(1) 
when Bob performs measurement just on the particle received from Alice
and (2) in the case when he utilizes the whole potential of the dense
coding, that is, he performs the measurement on the received particle and
the particle he had prior to the communication.
We also 
present a simple classical scenario
which might serve as a prototype of the dense coding. 
We generalize our results  also for qudits.
\end{abstract}
\pacs{03.67.-a,89.70.+c}
\begin{multicols}{2}
\section{Introduction}
Quantum dense coding is probably one of   
the most transparent demonstrations of the
power of quantum entanglement in quantum communication \cite{Nielsen}.
Bennett and Wiesner in their seminal paper \cite{BW}
have shown that entanglement shared between Alice
and Bob can increase a capacity of the quantum channel. Specifically,
it is well known that one-qubit channel 
can transmit at most one bit of information
\cite{Nielsen}. On the other hand, if the entanglement between 
two parties
is utilized, then uo to
 two bits of information can be transmitted via sending
just a single qubit from Alice to Bob: 
Let us suppose that Alice and Bob share a pair of two qubits prepared 
in the maximally entangled
state $|\psi\r_{AB}=|\psi^+\r=\frac{1}{\sqrt{2}}(|00\r_{AB}+|11\r_{AB})$.
Let Alice perform on her qubit one of the following four operations 
$\sigma_0=\openone,\sigma_1=\sigma_x,\sigma_2=\sigma_y,\sigma_3=\sigma_z$.
(here $\sigma_j$'s are Pauli matrices). 
In particular, the joint state $|\psi^+\r$
evolves according to Alice' s actions 
$\sigma_k\otimes\openone_B$ ($k=0,1,2,3$) into one of the
following  states
\be
\nonumber
|\psi^\pm\r&=&\frac{1}{\sqrt{2}}(|00\r\pm |11\r)\, , \\
|\phi^\pm\r&=&\frac{1}{\sqrt{2}}(|01\r\pm |10\r)\ . 
\ee
That is, Alice prepares one of the four mutually orthogonal states.
After that Alice sends her qubit to Bob. 
He performs the so-called {\it Bell measurement} on both qubits
to obtain one of the four possible outcomes associated with the operation
chosen by Alice. Thus, Alice and Bob can
communicate two bits of information per one usage of the channel.

In this scenario it is essential that Bob performs measurement on both
particles. The one he has received from Alice and the other which was
in his possession prior to 
 the communication via the channel. If Bob would perform
a measurement just on the particle he received from Alice, the amount of
information he gets is equal to zero. Certainly, if the two qubits were
not maximally entangled then the capacity of the channel is definitely less
than two. On the other hand, if Bob would perform a measurement only 
on the 
particle received from Alice he can get a non-zero information. 
What is interesting in this case is that Alice essentially encodes 
message into an unknown state $\rho_A = {\rm Tr_B}\rho_{AB}$ of her qubit
which might  or might not be entangled with Bob's qubit. 
The information is coded via the set of operations $U$ (see below). 
It is then the choice of Bob whether he boosts the capacity of the channel
by performing measurement on only Alice's particle or both particles.
In this paper we will analyze  the difference between these two scenarios. 
Specifically, we will assume the Alice's qubit to be in the state
$\varrho=\frac{1}{2}\openone+\vec{n}\cdot\vec{\sigma}$
and we evaluate a capacity of the noiseless channel for the case
when Bob performs measurement just on the  particle received from Alice 
and in the case when he utilizes the whole potential of the dense coding.
Comparing these two scenarios we will discuss the role of entanglement
for the dense coding and we will argue that not only entanglement but also
correlations are crucial for the dense coding. In other to illuminate
this argument in more detail we will present a simple classical scenario
which might serve as a prototype of dense coding. Finally we will generalize
our results for qudits.

\section{Capacity of noiseless qubit channels}
We start this section with a brief reminder of the definition of the
channel capacity
(see for example \cite{Presskill}).
Let $\pi_a$ represents the probability of the input state $\varrho_a$
of the system that will be transmitted via the quantum channel described by
the superoperator $\cE$. Since in this article we will deal only with
 {\em noiseless} channels the mapping $\cE$ corresponds to the identity, i.e.
$\varrho_{in}\to\varrho_{out}=\cE[\varrho_{in}]=\varrho_{in}$.
According to  Holevo 
\cite{Holevo98,Holevo98-2} the capacity of the channel is given 
by the following formula
\be
\label{holevo}
C(\cE)=\max_{\pi}\left[
S(\cE[\overline{\varrho}])-\sum_a\pi_a S(\cE[\varrho_a])
\right]
\ee
where $\overline{\varrho}:=\sum_a \pi_a\varrho_a$ is the {\it average state},
$S(\varrho)=-\T\varrho\log\varrho$ is the {\it von Neumann entropy} and 
the maximalization is taken over all possible input probabilities $\pi_a$.

Once the channel capacity is defined let us consider the first scenario:
Alice obtains a qubit which might or might not be a part of the
entangled pair. The qubit is prepared in the state
$\varrho=\frac{1}{2}\openone+\vec{n}\cdot\vec{\sigma}$
and Bob at the end of the communication channel performs measurement only
on the qubit received from Alice. She as a sender 
 is allowed to choose between unitary transformations 
$\U_a$ to encode the
message $a$ into the state $\varrho_a\equiv\vec{n}_\alpha
=\U_a\varrho\U_a^\dagger$. In what follows we will use a notation
$\varrho=\frac{1}{2}\openone+\vec{n}\cdot\vec{\sigma}\equiv\vec{n}$,
that is we will represent a state of a qubit by a vector $\vec{n}$ in a
three-dimensional space. The state space of a qubit corresponds to a Bloch
sphere of a unit radius. 
Since the encoding transformation is unitary it does not change 
the eigenvalues and consequently the entropy is preserved, 
i.e. $S(\vec{n}_a)=S(\vec{n})$. Therefore we obtain
\be
C(\varrho)=\max_{\pi}\left[S(\sum_a\pi_a\varrho_a)\right]-S(\varrho)\ .
\ee
Our aim is to maximize the entropy of the averaged state
$\sum_a\pi_a\varrho_a$.  
It is known, that the entropy achieves its maximum for the state
called as the {\it total mixture}, i.e. for the operator $\frac{1}{2}\openone$.
Thus the question is, whether it is possible to find such
a set of unitary transformations $\U_a$ for which
\be
\label{max_cond}
\sum_a \pi_a \U_a\varrho\U_a^\dagger=\frac{1}{2}\openone\ .
\ee

\begin{figure}
\includegraphics[width=9cm]{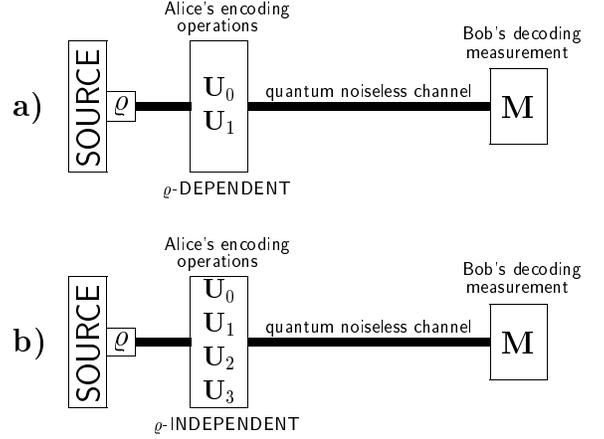}
\caption{A schematic description of a noiseless single qubit
channel with two different ways of coding.  Fig.~\ref{fig1}a
corresponds a situation when Alice is using just two coding
operations $U_0$ and $U_1$. In this case the capacity depends
on the input state $\varrho$. Fig.~\ref{fig1}b describes the
case when Alice is using four coding transformations. This codings
results in the situation when the capacity of the noiseless channel
does not depend on the input state $\varrho$.
}
\label{fig1}
\end{figure}

Let us first dicuss the case when $a=0$ or $1$
and  introduce the notation $\vec{n}_0=\vec{n},\vec{n}_1=\vec{m}$
(see Fig.~\ref{fig1}a)
The total mixture lies in the center of the Bloch sphere
representation of states of the qubit and corresponds to the vector
$\vec{0}$. Therefore, the condition (\ref{max_cond}) can be rewritten
as
\be
\vec{0}=\pi\vec{n}+(1-\pi)\vec{m}.
\ee
The convex sum of two vectors with equal lengths is equal to zero,
{\em if and only if} $\pi=1/2$ and $\vec{n}=-\vec{m}$, i.e. they have
opposite orientations. For pure states
it corresponds to the orthogonality of these states.   

As a result we get that in order to maximize the capacity, 
Alice needs to perform the unitary transformations 
$\U_0$ and $\U_1$ that generate two mutually orthogonal states, 
i.e. $\l\psi|\U_1\U_0|\psi\r=0\ $.
For a fixed (known) state $|\psi\r$ it is not a difficult task
but in a more general case 
(i.e. if the state $|\psi\r$ is unknown to Alice) it is impossible.
The transformation
$|\psi\r\to|\psi^\perp\r$ is  anti-unitary
 and therefore it cannot be performed perfectly 
(see \cite{Buzek}).
It means that in the case $a=0$ or $1$ it is impossible to
achieve $C(\varrho)=1-S(\varrho)$, if Alice does not know the
state $\varrho$ she gets. But is it entirely impossible? What
happens, if the number of applied unitaries $\U_a$
is larger than two? 

Let us choose the following four unitary transformations ($a=0,1,2,3$)
 $\U_0=\openone,\U_k=\vec{n}_k\cdot\vec{\sigma}$
for $k=1,2,3$ and $\vec{n}_k$ are three real three-dimensional vectors,
for which $\vec{n}_k\cdot\vec{n}_l=\delta_{kl}$ (see Fig.~ref{fig1}b) 
That is, 
the vectors $\vec{n}_k$ form a basis in 
the three-dimensional real vector space.
Let us put $\pi_a=1/4$ for all values of $a$.
Calculating the left hand side of Eq.(\ref{max_cond})
we obtain
\be
\nonumber
& &
\frac{1}{4}\sum_{a=0}^4\U_a\varrho\U_a^\dagger=\frac{1}{4}\left[\varrho+
\sum_k (\vec{n}_k\cdot\vec{\sigma})\varrho(\vec{n}_k\cdot\vec{\sigma})^\dagger\right]
\\
\nonumber
& & =\frac{1}{2}\openone+\frac{1}{4}\left[\vec{n}\cdot\vec{\sigma}+
\sum_k (\vec{n}_k\cdot\vec{\sigma})(\vec{n}\cdot\vec{\sigma})
(\vec{n}_k\cdot\vec{\sigma})\right] \\
\nonumber
& & =\frac{1}{2}\openone+\frac{1}{4}\left[
\vec{n}\cdot\vec{\sigma}+\sum_k (\vec{n}_k\cdot\vec{\sigma})
(\vec{n}\cdot\vec{n}_k\openone+i(\vec{n}\times\vec{n}_k)\cdot\vec{\sigma})
\right] \\
\nonumber
& & =\frac{1}{2}\openone+\frac{1}{4}\left[
\vec{n}\cdot\vec{\sigma}+\sum_k
\left((\vec{n}_k\cdot\vec{n})(\vec{n_k}\cdot\vec{\sigma})\right. \right.
\\
\nonumber
& & \ \ \ \ \ \ \ \ \ \ \ \ \ \ \ \ \ \ \ \ \ \ \ \ \ \ \ \ \ \ 
\left. \left.
-(\vec{n_k}\times(\vec{n}\times\vec{n}_k))\cdot\vec{\sigma}
\right)
\right]\\
\nonumber
& & =\frac{1}{2}\openone+\frac{1}{4}\left[2\sum_k(\vec{n}_k\cdot\vec{n})
(\vec{n}_k\cdot\vec{\sigma})-2(\vec{n}\cdot\vec{\sigma})\right]\\
\label{14}
& & =\frac{1}{2}\openone\, ,
\ee
where we used the following identities
\be
\nonumber
(\vec{n}\cdot\vec{\sigma})(\vec{m}\cdot\vec{\sigma})&=&(\vec{n}\cdot\vec{m})\openone+
i(\vec{n}\times\vec{m})\cdot\vec{\sigma}\, , \\
\nonumber
\vec{a}\times(\vec{b}\times\vec{c})&=&(\vec{a}\cdot\vec{c})\vec{b}-
(\vec{a}\cdot\vec{b})\vec{c}\, , \\
\nonumber
\sum_k(\vec{n}_k\cdot\vec{n})(\vec{n}_k\cdot\vec{\sigma})&=&
\sum_{k\alpha\beta} (n_k)^\alpha n^\alpha (n_k)^\beta\sigma_\beta =
\vec{n}\cdot\vec{\sigma}\,  , \\
\sum_k (n_k)^\alpha(n_k)^\beta &=& \delta^{\alpha\beta}\ \ 
{\rm (completness)}\, ,
\ee
and $(n_k)^\alpha$ denotes the $\alpha$-th component of the vector $\vec{n}_k$.

What we have shown here is that 
if Alice uses the four unitary operations $\U_0=\openone,
\U_k=\vec{n}_k\cdot\vec{\sigma}$ and the information 
source produces 
messages $a=0,1,2,3$ with equal probabilities, (i.e. $\pi_a=1/4$),
then Eq.(\ref{max_cond}) holds for a general (unknown) state $\varrho$.
Therefore, the capacity of the noiseless qubit 
channel is given by 
\be
\label{normalcap}
C(\varrho)=1-S(\varrho)\ .
\ee
The set of unitaries is universal in a sense that 
their choice is independent of the state $\varrho$. It is interesting
that the mentioned universality cannot be obtained by using
only two-valued encoding $\U_a$, but with four-valued encoding
it is possible. We stress once again that in this scenario the entanglement
has not been employed at all. For that reason we will denote the capacity
(\ref{normalcap}) as $C^{normal}$.

\section{Dense coding with partially entangled states}
As we have shown in the Introduction, 
the dense coding protocol is based on a very specific property
of  maximally entangled states. Namely, it is based on the possibility 
to generate the basis of maximally entangled states just by
local unitary operations realized by Alice. There are four such operations
which generate four mutually orthogonal states.
This property  is no longer valid,
if the qubits are entangled only partially.

In this section we will consider a situation when Alice and Bob
share a partially entangled pair of quits such that Alice's qubit is 
in the state
$\varrho=\frac{1}{2}\openone+\vec{n}\cdot\vec{\sigma}\equiv\vec{n}$.
(We note that in the
case of maximally entangled pair
Alice's qubit is in the maximally mixed state $\frac{1}{2}\openone$.) 
Bob is going to utilize the dense coding strategy, that is he
will perform a measurement on both qubits. 
The question is which type of operations Alice has to perform in order
to minimalize the mutual overlap between
the states $\varrho_a$. For these operations we can expect the maximal
capacity of the quantum channel.
This question has been addressed in Ref.~\cite{Werner,Plenio,Ziman}. 
Here we just
briefly evaluate the capacity
for noiseless one-qubit channel. 
Specifically, let us consider that Alice 
realizes one of the four unitary transformations
$\U_a$ to obtain four states 
$\varrho_a=(\U_a\otimes\openone)\varrho_{AB}(\U_a^\dagger\otimes\openone)$
of a two-qubit system. 
Since again the transformations are unitary, it follows
that the second term in the expression (\ref{holevo})
for the capacity equals the entropy of the joint state, i.e.
$S(\varrho_{AB})$. We can write
\be
C(\varrho_{AB})=\max_\pi\left[S(\sum_a\pi_a\varrho_a)\right]-S(\varrho_{AB})\ .
\ee
The question remains the same as before. What is the maximal value 
of the first term?
Since the whole Hilbert space is four-dimensional, the largest possible
value of the entropy
is $\log 4=2$. However, is it possible to
achieve this value, if the unitary transformations
must have the form of $\U_a\otimes\openone$? 

Firstly consider the direct generalization
of the Bennett and Wiesner example, that is, let Alice
performs just four possible transformations $\U_a\otimes\openone$.
The general state of a two qubit system can be uniquely expressed in the
following way
\be
\varrho_{AB}=
\varrho_A\otimes\varrho_B+
\sum_{cd}\gamma_{cd}\sigma_c\otimes\sigma_d\, ,
\ee
where $\varrho_A=\T_B\varrho_{AB}$ and 
$\varrho_B=\T_A\varrho_{AB}$ are the reduced density operators
describing  states of the subsystems (Alice and Bob).

If we again set the probabilities $\pi_a=1/4$, then
the average state
\be
\overline{\varrho}_{AB}&=&\frac{1}{4}\sum_a (\U_a\otimes\openone)\varrho_{AB}
(\U_a^\dagger\otimes\openone)
\ee
can be calculated. In Ref.~\cite{Ziman} it has been shown that 
if we require that  the four Alice's operations $U_a$ are
independent of  $\varrho_{AB}$ 
then these unitaries must have the form $\U_a\otimes\openone$
with  $\U_0=\openone,\U_k=\vec{n}_k\cdot\vec{\sigma}$. That is, these
operations are exactly the same as those derived in previous section
in a completely different scenario.
It means the vectors $\vec{n}_k$ form an
 orthonormal basis in the three-dimensional
real vector space.

Using the previous results (Sec.II) we can write
\be
\overline{\varrho}_{AB}=\frac{1}{2}\openone\otimes\varrho_B+
\sum_{cd}\gamma_{cd}\left(\frac{1}{4}
\sum_a\U_a\sigma_c\U_a^\dagger\right)\otimes\sigma_d\ .
\ee
If we insert the sigma operator $\sigma_c$ (instead of
the state $\varrho$) into the calculation (\ref{14}) it 
gives us  
\be
\frac{1}{4}\sum_a \U_a\sigma_c\U_a^\dagger=0\ .
\ee
Thus, we find that
\be
\overline{\varrho}=\frac{1}{2}\openone\otimes\varrho_B\, ,
\ee
and for the dense coding capacity we obtain the formula
\be
\nonumber
C(\varrho_{AB})&=&S(\frac{1}{2}\openone\otimes\varrho_B)-S(\varrho_{AB})\\
\label{densecap}
&=&1+S(\varrho_B)-S(\varrho_{AB})\ .
\ee
In the last equality we used the following property 
of the entropy function  
$S(\varrho\otimes\xi)=S(\varrho)+S(\xi)$. 

We conclude this section with two remarks:
\leftline{\bf Remark 1. {\it Maximal capacity}}
We still did not solve the original question of maximalizing
the capacity in its full generality. 
It is an open problem, 
whether we can raise the capacity by applying more unitaries, like
it was done in Section II. 
We will get back to this problem at the end of this paper.
\\
\leftline{\bf Remark 2. {\it The asymmetry of the dense coding}}
Let us note, that the obtained capacity of the noiseless qubit channel using
the dense coding strategy is not symmetric with respect to
the exchange of Bob and Alice. Suppose the same situation as before,
i.e. Alice and Bob share a pair of qubits in a state $\varrho_{AB}$.
If Bob send the messages to Alice (using dense coding strategy) 
we obtain
\be
\nonumber
C_{B\to A}&=&1+S(\varrho_A)-S(\varrho_{AB})\\
&\ne & 1+S(\varrho_B)-
S(\varrho_{AB})=C_{A\to B}\, , 
\ee
since in general $S(\varrho_A)\ne S(\varrho_B)$. In fact, there is no reason 
to expect the equality, since Bob and Alice use different signal states.\\

Finally, in what follows we will denote the 
 capacity (\ref{densecap}) related to dense coding as $C^{dense}$.

\section{Correlations are crucial}
\begin{figure}
\includegraphics[width=9cm]{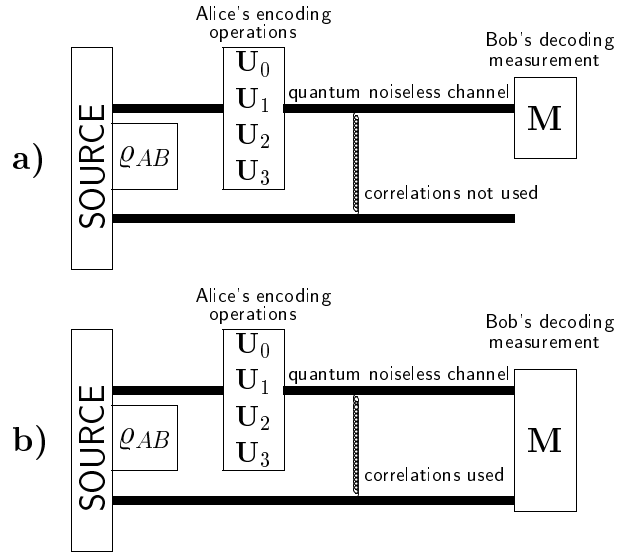}
\caption{Schematic description of the communication between Alice and Bob.
Fig.~\ref{fig2}a describes a situation when Alice and Bob
share a correlated pair of qubits but Bob does perform a measurement
only on the particle received from Alice. In this case he does not
utilize the potential of the dense-coding strategy. In Fig.~\ref{fig2}b
Bob performs measurement on both particles - the one that he received prior
to the communication directly from the source of correlated particles and
the particle received from Alice. In this second case correlations between
particles are used to enhance the capacity of the noiseless channel.
}
\label{fig2}
\end{figure}

To compare the capacity of the 
dense coding scenario (\ref{densecap}) with the capacity of 
 the noiseless qubit channel without using the dense coding
strategy (\ref{normalcap}), we see that always $C^{dense}\ge C^{normal}$. 
Both of these scenarios use the same unitary transformations realized
on the same state
$\varrho=\frac{1}{2}\openone+\vec{n}\cdot\vec{\sigma}\equiv\vec{n}$.
of Alice's qubit 
prepared prior the communication. 
The four unitary transformation are used  to
generate the input signals. 
The choice of the scheme to be used depends on 
Alice and Bob. In fact, it is Bob's ``free will'' whether he uses
the second qubit and thus whether he establishes the dense coding
communication or not (see Fig.~\ref{fig2}). 
 In some sense Alice does not have 
to do anything different in either case. 
She simply always chooses one of the four possible
unitary transformations.
In particular, the difference between channel capacities associated with
these two strategies
\be
\nonumber
C^{dense}_{A\to B}-C^{normal}_{A\to B}&=&
C^{dense}_{B\to A}-C^{normal}_{B\to A}\\ 
\nonumber
&=&
S(\varrho_A)+S(\varrho_B)-S(\varrho_{AB})\\ &=& C_\varrho(A,B) \,  ,
\label{4.1}
\ee
allows us to make the conclusion 
 that it is  not exclusively the entanglement, but the
correlations {\em per se} which 
are crucial in the dense coding scenario.
We remind us that the function denoted as $C_\varrho(A,B)$ 
in Eq.~(\ref{4.1}) is  the correlation function,
or to the so-called {\it quantum mutual information}
\cite{Nielsen}.
In order to appreciate this result we will consider a simple example
of a dense coding within a classical context.

\subsection{ Dense coding in classical settings}
Let us consider the following {\em classical} scenario: 
Alice and Bob share the maximally correlated classical 
state of two two-level particles (bits). 
It means that the source 
produces two particles described by 
the joint classical state
$p(00)=p(11)=1/2$ and $p(01)=p(10)=0$. One of them is sent to Bob
and the other to Alice. 
In the classical case (unlike quantum) Alice can perform only
two operations on her bit. 
If she wants to send the message 0, she does nothing, and if the message
is 1, then she performs classical $NOT$ on her bit. That is, if she obtains
a bit with the value $j$, 
then she sends the bit with the value $j\oplus k$, where $k$
is the message she wants to send. After Bob receives her bit, he can
decide to measure it with no reference to his bit what-so-ever. In this
case he obtains no information and capacity of the channel is zero
(which is an analogue of the quantum case when Alice is performing
operations on the qubit in 
 the state $\frac{1}{2}\openone$ while Bob measures only this qubit).
On the other hand, when
Bob receives Alice's bit he can compare
it with the original one he obtained from the source. If their values 
coincide, then he knows that Alice sent him the message 0. 
If he finds a 
difference, then he knows that Alice realized the $NOT$ operation 
and the received message is 1. Thus, Alice and Bob can communicate 
one bit of information using the ``dense coding'' strategy. 
In this classical picture it is more
illustrative that Bob's bit plays a role of a secure key. Formally,
the classical and quantum situations are the same. 
The classical dense coding (with the maximally correlated state)
is completely equivalent to the communication, where
Alice and Bob share a secure key. They need two bits
(as classical physical systems) to transmit one bit of information.
But the transmission is as secure, as in the quantum case,
only the existence of the potential eavesdropper is not detectable. 
(On the other hand the knowledge about the value
of the bit transmitted via the channel is useless for the eavesdropper).

\subsection{Entanglement measure}
We might consider to utilize
the dense coding as a way how 
to define an entanglement measure 
(see Refs.~\cite{Plenio,Ziman}). 
From our previous results if follows that not only
the entanglement but  the
correlations between Alice and Bob
are essential for the dense coding. So the question is whether
one can use the phenomenon to quantify the degree of entanglement between
Alice and Bob.

The main problem in this context is that there might be several 
views how to define what the dense coding is. 
Within the context of quantum information processing the 
classical-quantum analogies are not obtained 
as a consequence of some rigorously defined procedures,
but rather
they are based on  vague (intuitive) mathematical similarities.
If one represents the dense coding as any strategy that breaks the
limit on the capacity of the one-qubit channel, then one can say that
whenever the capacity $C(\varrho_{AB})>1$ then 
the state $\varrho_{AB}$
is entangled. Obviously this is right, since from Eq.~(\ref{densecap}) 
it follows that 
either $S(\varrho_A)>S(\varrho_{AB})$, or $S(\varrho_B)>S(\varrho_{AB})$.
On the other hand, 
it is known, however, that not all entangled states have such property.
We have tried to define the entanglement as the difference
between the ``normal'' and ``dense'' capacities, but, of course,
such definition strongly depends on the definition what 
the ``normal'' strategy is.
We did it in a simple and natural way, but it might be that 
other definitions could bring some new insights into the
problem of the entanglement. Our choice (at least)
enables us to find the classical analogue of the dense coding strategy.

Probably it is worth to note that 
there exists a formal mathematical
relation between the difference of these capacities and
the entanglement of formation $E_F$ (see below).
Let us consider Eq.~(\ref{4.1}) describing the difference between
$C^{dense}$ and $C^{normal}$. This difference is equal to the mutual
information $C_\varrho(A,B)$. Based on this expression we can introduce
a function $E(A,B)$ defined as 
\be
\label{def}
E(\varrho_{AB})&=&
\min_{p_k,|\psi_k\r_{AB}}\left[
\sum_k p_k C_{|\psi_k\r}(A,B)\right]\\
\nonumber
&=&
\min_{p_k,|\psi_k\r_{AB}}\left[
\sum_k p_k (S(\varrho_A^k)+S(\varrho_B^k))
\right]
\ee   
where the minimum is taken over all convex decompositions
of the state $\varrho_{AB}$ into pure states and
$\varrho_A^k=\T_B|\psi_k\r_{AB}\l\psi_k|$,
$\varrho_B^k=\T_A|\psi_k\r_{AB}\l\psi_k|$.
Since for pure states $|\psi\r_{AB}$ the entropies
of the subsystems are the same, $S(\varrho_A)=S(\varrho_B)$, it follows
that the function $E(\varrho_{AB})$ is proportional
to the {\it entanglement of formation} $E_F$, i.e.
\be
E(\varrho_{AB})=
\min_{p_k,|\psi_k\r_{AB}}2 \left[
\sum_k p_k S(\varrho^k_A)
\right] = 2 E_F(\varrho_{AB})\ .
\ee 
Of course, the meaning of the last equality
is rather vague. We do not have any compelling reasons why to use the 
definition given in Eq.(\ref{def}).
We can only  argue that
it excludes the possibility to substitute the source of
the pairs of qubits by  local sources that are allowed
to communicate via classical channels.
In other words, when the source can be replaced
by two local sources connected by a classical communication line, 
the entanglement $E(\varrho_{AB})$ vanishes.

\section{Instead of conclusions: Noiseless qudit channel}

In this paper we have shown that quantum dense coding has its origin in
correlations between two parties rather than simply in entanglement.
In order to justify our argument we have considered a situation 
that 
Alice has a  qubit  in the state
$\varrho=\frac{1}{2}\openone+\vec{n}\cdot\vec{\sigma}$
and we evaluated a capacity of the noiseless channel for two cases: 
(1) 
when Bob performs measurement just on the particle received from Alice
and (2) in the case when he utilizes the whole potential of the dense
coding, that is, he performs the measurement of the particle received 
from Alice  and
the particle he had prior the communication.
We have also
presented a simple classical scenario
which might serve as a prototype of dense coding. 
In all our discussions we have considered that Alice and Bob share a pair
of qubits. In a conclusion we show that our results are valid also for
qudits.

By qudits we understand  $d$-dimensional quantum objects. 
The main property we have used in our
discussion with qubits is expressed by Eq.(\ref{max_cond}) and
for qudits this expression takes the following form
\be
\label{x}
\sum_a\pi_a\U_a\varrho\U_a^\dagger
=\frac{1}{d}\openone=\overline{\varrho}\ .
\ee
We can express the general qudit state in the form
$\varrho=\frac{1}{d}\openone+\vec{n}\cdot\vec{\Lambda}$,
where $\vec{\Lambda}=(\Lambda_1,\dots,\Lambda_{d^2-1})$
is the set of $d^2-1$ Hermitian traceless operators,
for which $\T\Lambda_\alpha\Lambda_\beta=d\delta_{\alpha\beta}$
with $\alpha,\beta=1,\dots,d^2-1$. 
Let us choose the set of $d^2$ unitary operators $\U_a$, for which
the similar property holds, i.e. $\T\U_a^\dagger\U_b =d\delta_{ab}$,
but $a,b=0,1,\dots,d^2-1$. We assume that $\pi_a=1/d^2$.
Introducing this notation we can rewrite the above equation as 
\be
\label{xx}
\frac{1}{d^2}\sum_a\U_a\varrho\U_a^\dagger=\frac{1}{d}\openone+
\frac{1}{d^2}\sum_{\alpha=1}^{d^2-1}n_\alpha\left(
\sum_{a=0}^{d^2-1}\U_a\Lambda_\alpha\U_a^\dagger\right)\, .
\ee
Next, we will show that the second term in the right-hand side of
Eq.~(\ref{xx}) vanishes. Let 
$|\psi\r=\sum_k\frac{1}{\sqrt{d}}|k\r \otimes|k\r $
be the maximally entangled state of two qudits and let us calculate
the mean value of the operator $\xi_\alpha\otimes\openone$, where 
$\xi_\alpha:=\sum_a\U_a\Lambda_\alpha\U_a^\dagger$. That is
\be
\nonumber
\l\psi|\xi_\alpha\otimes\openone|\psi\r
&=&\sum_a \l\phi_a|\Lambda_\alpha\otimes\openone|\phi_a\r 
\\
&=&\T\Lambda_\alpha\otimes\openone=\T_1\Lambda_\alpha\T_2\openone=0\, , 
\ee
where we used the notation $|\phi_a\r:=\U_a^\dagger\otimes\openone
|\psi\r $ and 
the identity $\l\phi_a|\phi_b\r=\frac{1}{d}\T\U^\dagger_a\U_b=\delta_{ab}$
which 
implies that the vectors $|\phi_a\r$ form an orthonormal basis
in the Hilbert space of two qudits. 
The last equality is the consequence of the tracelessness of $\Lambda_\alpha$.
Since the mean value 
$\l\psi|\xi_\alpha\otimes\openone|\psi\r$ 
for all states $|\psi\r$ equals to zero, 
we can conclude that the operator $\xi_\alpha\otimes\openone$ vanishes
as well as the operator $\xi_\alpha$ for all $\alpha$. 
As a result we find that the second term in (\ref{xx})vanishes and 
the condition (\ref{x}) holds.
We proved the property that enables us to generalize our previous
results. It is easy to see
that for the capacities of noiseless qudit channels we get
\be
\nonumber
C_{A\to B}^{normal}(\varrho_A)&=&\log_2 d - S(\varrho_A)\, , \\
C^{dense}_{A\to B}(\varrho_{AB})&=&\log_2 d+S(\varrho_B)-S(\varrho_{AB})
\, , 
\ee
where we used the fact that 
$\max_\pi\left[S(\overline{\varrho})\right]=\log_2 d$
is achievable, since we showed that $\overline{\varrho}=\frac{1}{d}\openone$.
Our generalization also answers the question stated at the end of 
Section II (see Remark 1). Since it is impossible to find the set
of $d^2$ local unitaries  of the form $\U\otimes\openone$
satisfying the orthogonality condition
$\T\U_a^\dagger\U_b=d\delta_{ab}$, it follows that the 
dense coding capacity cannot achieve the value 
$C(\varrho_{AB})=\log_2 d^2 - S(\varrho_{AB})$. That  is,
the maximal capacity cannot be achieved due to the fact
that by applying the encoding transformations $\U\otimes\openone$
we cannot obtain the averaged state of the form
$\overline{\varrho}_{AB}= \frac{1}{d^2}\openone_{AB}$ (for the
general initial state $\varrho_{AB}$). The best encoding transformations
which lead to a maximum capacity generate the averaged state 
$\overline{\varrho}_{AB}$ of the form 
$\overline{\varrho}_{AB}=\frac{1}{d}\openone_A\otimes\varrho_B$.

\acknowledgements
This work was supported in part
by the European Union  projects EQUIP (IST-1999-11053) and QUBITS
(IST-1999-13021).


\end{multicols}
\end{document}